\documentclass[aps,prb,amsmath,amssymb,twocolumn,groupedaddress,showpacs]{revtex4-1}
\usepackage{graphicx}

\begin{document}
	
	\title{Spaser Based on Graphene Capillary}
	
	\author{Sadreddin Behjati Ardakani}
	\email[]{behjati@ee.sharif.ir}
	\affiliation{Department of Electrical Engineering, Sharif University of Technology, Azadi Avenue, Tehran, Iran}
	
	\author{Rahim Faez}
	\email[]{faez@sharif.ir}
	\affiliation{Department of Electrical Engineering, Sharif University of Technology, Azadi Avenue, Tehran, Iran}
	
	\date{\today}
	
\begin{abstract}
	In this paper, we propose a structure for graphene spaser and develop an electrostatic model for quantizing plasmonic modes. Using this model, one can analyze any spaser consisting of graphene in the electrostatic regime. The proposed structure is investigated analytically and the spasing condition is derived. We show that spasing can occur in some frequencies where the Quality factor of plasmonic modes is higher than some special minimum value. Finally, an algorithmic design procedure is proposed, by which one can design the structure for a given frequency. As an example, a spaser with plasmon energy of 0.1\,eV is designed.
\end{abstract}


\maketitle

\section{Introduction}
\label{sec:int}

Spaser (Surface Plasmon Amplification by Stimulated Emission of Radiation), as its name suggests, going to be a counterpart of laser in sub-wavelength dimensions. The diference between spaser and laser is that laser emits photons but spaser emits intense coherent surface plasmons (SPs). The idea was emerged after trying to overcome the main shortcoming of laser. Because laser emits photons, it must suffer from photon's diffraction limit. The electromagnetic field of photons cannot be concentrated in spots which are qualitatively smaller than half their wavelength. This is a fundamental theoretical limit and so cannot be circumvented. So spaser inventors, Stockman and Bergman, suggested using another particle, instead of a photon, which does not have this theoretical constraint \cite{bergman2003surface}. Their idea was to utilize the extra confine nature of SPs. SPs can confine in regions much smaller than their wavelengths. In 2003, Stockman and Bergman published the first paper about spaser and introduced the word spaser to the literature \cite{bergman2003surface}. Since that, many people and groups focused on analyzing and realizing it. In 2009, Noginov \textit{et al.} demonstrated an experimental spaser using an aqueous solution of gold nanoparticles surrounded by dye-doped silica shell as a gain medium \cite{noginov2009demonstration}. In 2010, Stockman proposed a plasmon amplifier using spaser, and analyzed its equation of motion using optical Bloch equations. The author claimed that the spaser could not be analyzed classically \cite{stockman2010spaser}. Zhong and Li tried to analyze the spaser semi-classically in 2013 \cite{zhong2013all}. Dorfman \textit{et al.} focused on the full quantum mechanical description of spaser in 2013 \cite{dorfman2013quantum}. In 2014, Apalkov \textit{et al.} proposed a graphene-based spaser \cite{apalkov2014proposed}. Until now many papers have been published, covering many aspects of spaser \cite{andrianov2011forced, khurgin2012injection, li2013electric, parfenyev2014quantum, rupasinghe2014spaser, jayasekara2015multimode, totero2016energy}.

Spaser, similar to its partner, laser, consists of two main parts: a medium for supporting SP modes, and an active or gain medium. SPs can propagate along interface between two materials which one of them has negative dielectric constant. Metals have negative permittivities below their plasma frequencies, and thus a majority of papers focus on them as a medium for supporting and propagating plasmon modes. But metals are not the ideal ones. Metal losses avoid plasmons to propagate along long distances. In this paper, we will use graphene instead of metal.

Graphene is a material which forms by a 2D arrangement of carbon atoms in a honeycomb lattice bonding by strong $\mathrm{sp}^2$ hybridized covalent $\sigma$ bonds \cite{novoselov2005two}. The $p_z$ electrons of carbons, lying in $\pi$ orbitals, give the graphene some extraordinary electronic properties. The graphene electrons, near Dirac points, have a linear dispersion, so behave like massless Dirac fermions. Plasmons can propagate along and confine close to graphene about an order of magnitude more stronger than metals.

The active medium provides the energy required for initiating and maintaining the spasing process. The main factor for choosing active medium is pumping mechanism. Similar to laser, pumping method can be optical, chemical, electrical, and so on. In our research, we are going to use electrical pumping method by utilizing a Quantum Wire (QW) as gain medium.

In this paper, the full quantum mechanical approach is used for analyzing the structure. The most important quantity in quantum mechanics is the system's Hamiltonian. The Hamiltonian of the entire system is $H=H_\mathrm{sp}+H_\mathrm{am}+H_\mathrm{int}$,
where $H_\mathrm{sp}$, $H_\mathrm{am}$, and $H_\mathrm{int}$ are SP, active medium, and interaction Hamiltonians, respectively. The individual Hamiltonian parts are quantized in subsequent sections.

According to the aforementioned discussion, the paper is organized as follows: In section \ref{sec:structure}, we introduce our proposed structure which will be used throughout the paper. Section \ref{sec:SPHam} is devoted to the Hamiltonian of SP field and quantizing it. Section \ref{sec:gainHam} concentrates on active medium, and section \ref{sec:intHam} is dedicated to the interaction mechanism and derivation of the spasing condition. Moreover, in this section, we suggest a procedure for designing the structure.

\section{The Main Structure}
\label{sec:structure}
\begin{figure}[tbp]
\centering
\includegraphics[width=0.8\linewidth]{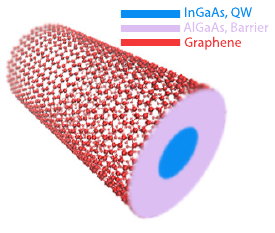}
\caption{(Color online) The proposed structure. The materials which are included in the structure are distinguished by different colors. The shown structure is embedded in a matrix of InP.}
\label{fig:structure}
\end{figure}

Our proposed structure consists of a graphene-coated cylindrical layered semiconductor heterostructure with average dielectric constant $\epsilon_1$ as shown in Fig.~\ref{fig:structure}. The semiconductor heterostructure makes up of two layers of semiconductors with different energy gaps. The energy gap of inner rod is lower than outer shell one, so that the heterostructure forms a QW system. The inner rod plays the role of QW and the outer shell is its barrier. The graphene-coated system is embedded in a matrix with dielectric constant $\epsilon_2$. Regarding the extra-confine nature of SPs, it can be assumed that $\epsilon_1$ is equal to outer shell dielectric constant, because, roughly speaking, the inner rod only sense the weak tail of SPs' field.
For numerical calculations, the specific material system, $\mathrm{Al}_{0.48}\mathrm{In}_{0.52}\mathrm{As}/\mathrm{Ga}_{0.47}\mathrm{In}_{0.53}\mathrm{As}$, is used. All the materials are chosen such that they are lattice matched to the matrix, InP, at room temperature 295\,K. The required physical parameters of these materials are tabulated in Table~\ref{tab:param}.
\begin{table}[tb]
\renewcommand{\arraystretch}{1.3}
\caption{\label{tab:param}\bf Physical parameters of materials which are used in this paper. All the alloys are chosen such that to be lattice matched with InP at 295\,K. Dielectric constants of ternary alloys are calculated from interpolation method \cite{chuang2009physics,vurgaftman2001band}. In this table, $\epsilon_r$, $m^*$, and $m_0$ represent dielectric constant, electron's effective mass, and electron mass, respectively.}
\begin{tabular*}{\linewidth}{@{\extracolsep{\fill}}lcc}
\hline\hline
Material & $\epsilon_r$ & $m^*/m_0$\\
\hline
$\mathrm{InP}$ & 12.56 & 0.077\\
$\mathrm{Al}_{0.48}\mathrm{In}_{0.52}\mathrm{As}$ & 12.46 & 0.075\\
$\mathrm{Ga}_{0.47}\mathrm{In}_{0.53}\mathrm{As}$ & 13.60 & 0.041\\
\hline\hline
\end{tabular*}
\end{table}

The graphene cylinder will support plasmonic modes. The cylindrical symmetry of this configuration makes possible deriving plasmonic modes analytically and also reducing its geometrical parameters to only one, tube radius $a$. So one can tune the plasmons by adjusting fewer parameters.

The QW is used as an active medium to provide energy for plasmons to maintain spasing. By applying electric potential difference between graphene and QW, the electrons in QW excite. Depending on degree of coupling strength between QW and SPs, the energy can interchange among electrons and SPs. The oscillation of energy exchange can continue steadily under some conditions. Next sections deal with finding this condition.

Throughout this paper, the cylinder is assumed to be infinitely long such that the edge effects can be neglected, and also its radius to be large enough such that size quantization effects do not influence its conductivity, significantly.

\section{\label{sec:SPHam}SP Hamiltonian}
For quantizing the SP Hamiltonian, The orthogonal potential modes of the structure should be derived. So, this section is divided into two subsections. The first subsection deals with extracting the potential modes and the second one is about writing SP Hamiltonian in quantized form.
 
\subsection{Graphene Cylinder Plasmonic Modes}
\label{subsec:modes}
For calculation purposes we need conductivity, so before proceeding further, the graphene conductivity is introduced. Conductivity of graphene is written as, $\sigma(\omega)=\sigma_\mathrm{intra}(\omega)+\sigma_\mathrm{inter}(\omega)$, where \begin{equation}
\sigma_\mathrm{intra}=\frac{2e^2k_BT}{\pi\hbar^2}\frac{i}{\omega+i\tau^{-1}}\ln\left[2\cosh\left(\frac{E_F}{2k_BT}\right)\right],
\end{equation}
and
\begin{eqnarray}
&&\hspace{-1.5cm}\sigma_\mathrm{inter}=\frac{e^2}{4\hbar}\times\nonumber\\
&&\hspace{-1.3cm}\left(\mathrm{H}(\omega/2)+\frac{4i(\omega+i\tau^{-1})}{\pi}\int_0^\infty\frac{[\mathrm{H}(\epsilon)-\mathrm{H}(\omega/2)]\,\mathrm{d}\epsilon}{(\omega+i\tau^{-1})^2-4\epsilon^2}\right),
\end{eqnarray}
with the following definition,
\begin{equation}
\mathrm{H}(\epsilon)=\frac{\sinh(\hbar\epsilon/k_BT)}{\cosh(E_F/k_BT)+\cosh(\hbar\epsilon/k_BT)}.
\end{equation}
In the above relations $e$, $k_B$ and $\hbar$ are elementary charge, Boltzmann  and reduced Planck constants, respectively. $T$ and $\tau\simeq0.4\mathrm{ps}$ \cite{apalkov2014proposed} are temperature and electron relaxation time, respectively. In the conditions where $\hbar\omega<2E_F$ and $\hbar\omega<\hbar\omega_\mathrm{oph}$, where $\hbar\omega_\mathrm{oph}\simeq0.2\,\mathrm{eV}$ is optical phonon energy in graphene, Drude-like profile is a good approximation for graphene's conductivity \cite{jablan2009plasmonics},
\begin{equation}
\sigma=\frac{e^2E_F}{\pi\hbar^2}\frac{i}{\omega+i\tau^{-1}}.\label{eq:drude}
\end{equation}

After this short introduction to graphene's conductivity, we turn back to the main target, which is mode calculation. The quasi-electrostatic approximation is utilized throughout the paper. This assumption is plausible due to very confine nature of plasmon modes \cite{christensen2017classical}.

Considering cylindrical symmetry of the structure, one can guess the following electric potential profile,
\begin{equation}
\phi(\mathbf{r},t)=\left\lbrace\begin{array}{ll}
A_m\mathrm{I}_m(k\rho)\exp i(kz+m\varphi-\omega_{k,m}t) &\rho\le a\\
B_m\mathrm{K}_m(k\rho)\exp i(kz+m\varphi-\omega_{k,m}t) &\rho>a,
\end{array}\right.
\end{equation}
where $A_m$ and $B_m$ are dependent arbitrary coefficients to be determined; $k$, $m$, and $\omega_{k,m}$ are mode indices, and corresponding frequencies, respectively. $\mathrm{I}_m$ and $\mathrm{K}_m$ are $m$'th order modified Bessel functions of first and second kind, respectively. In this paper, $\rho$, $\phi$, and $z$ symbols are reserved for radial and angular coordinates in cylindrical coordinate system. The unit vector along any direction is denoted by adding a hat symbol above the vector associated with that direction, moreover the hat symbol is reused for representing operators in the last section, without adding any ambiguity. Furthermore, without any loss of generality, it is assumed that graphene cylinder is oriented along the $z$ direction.

\begin{figure}[tb]
\centering
\includegraphics[width=\linewidth]{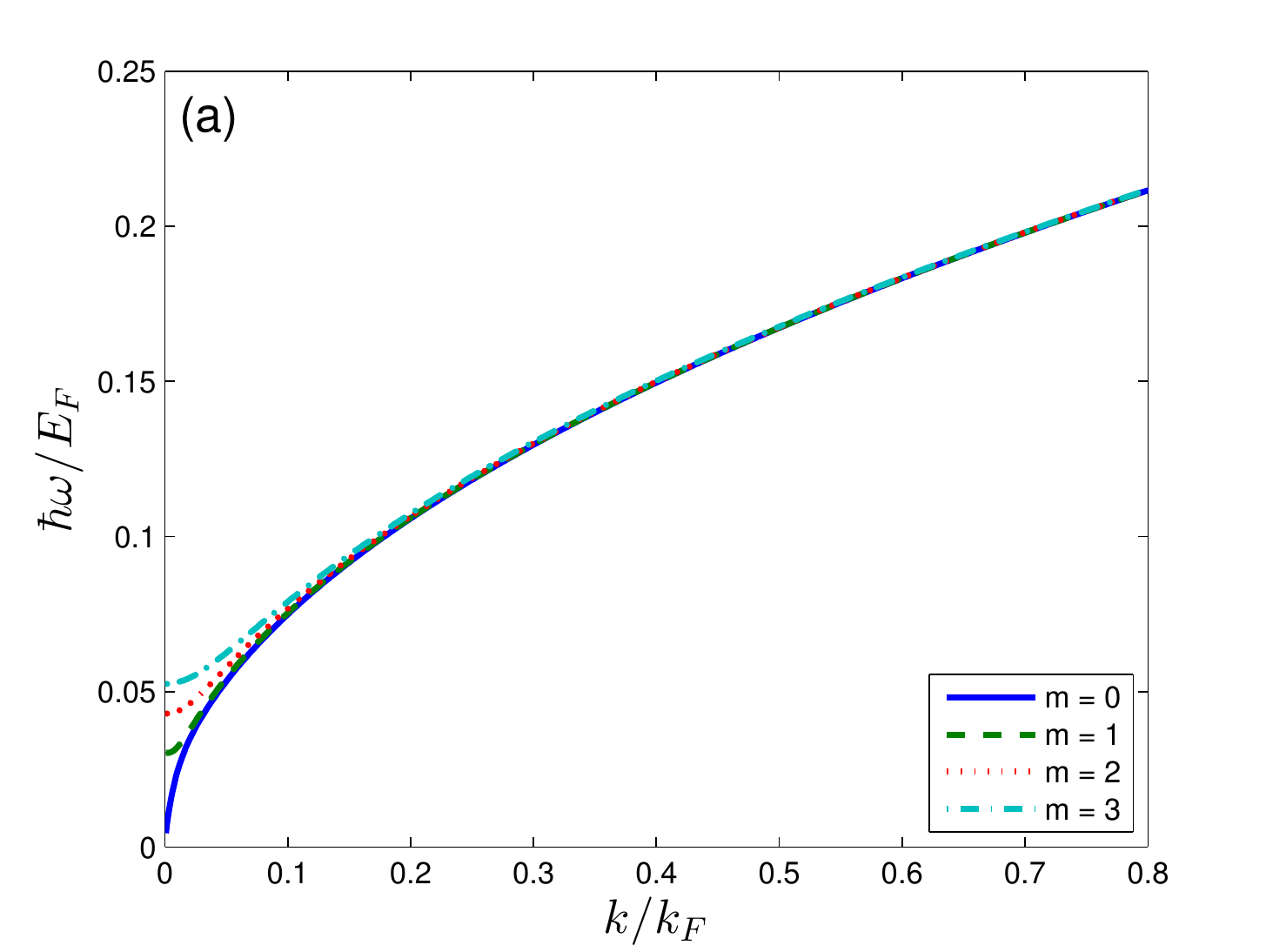}
\includegraphics[width=\linewidth]{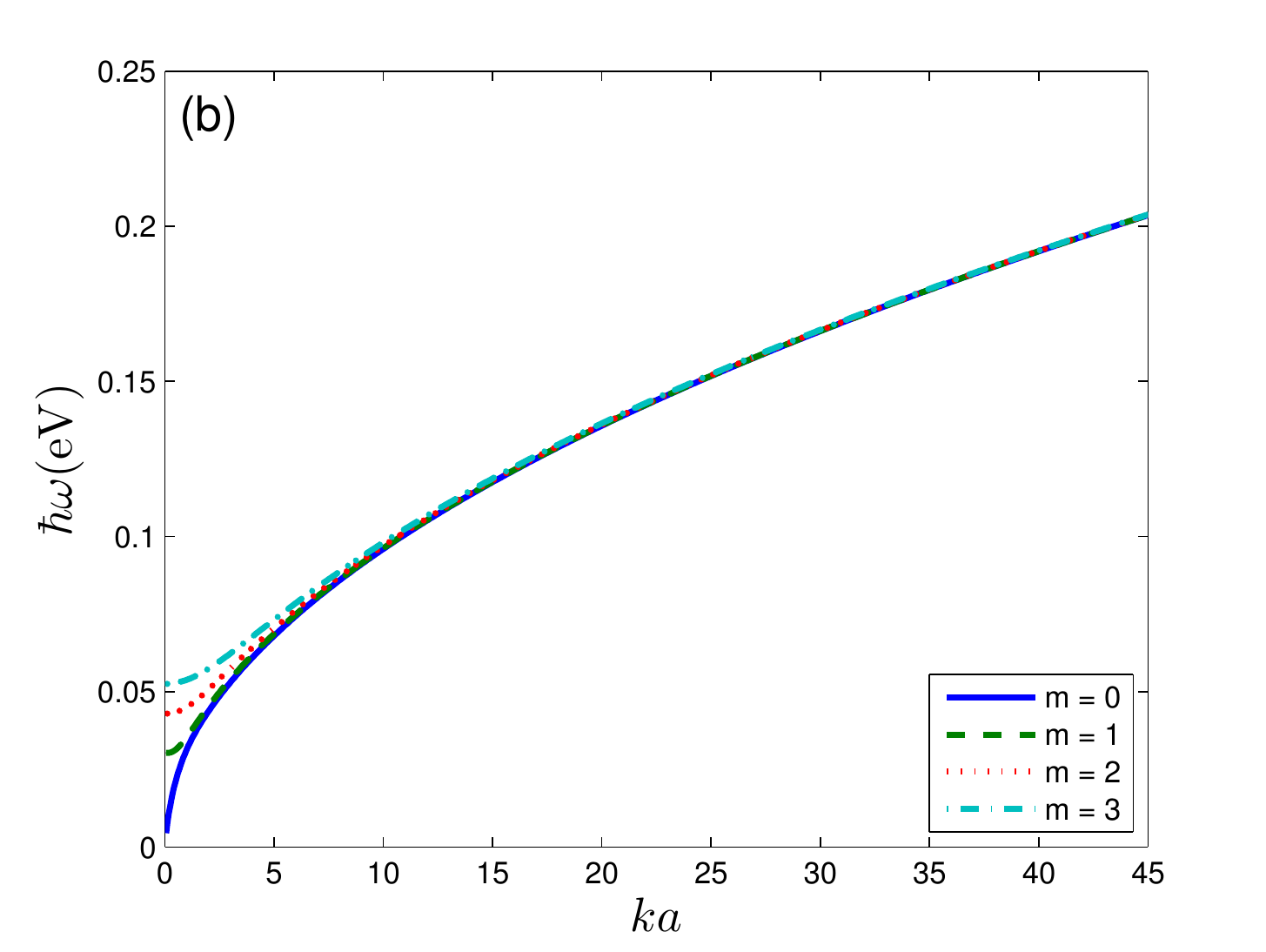}
\caption{(Color online) Plasmons dispersion for different Fermi energies. In deriving these curves The Drude approximation is assumed. (a) Normalized plasmon energy as a function of normalized wavenumber. (b) Plasmon energy versus dimensionless $ka$. These figures show that for large enough wavenumber all the modes become degenerate.}
\label{fig:disp}
\end{figure}

For this guess to be a valid solution, it must fulfill boundary conditions. After application of potential continuity across the boundary, $\rho=a$, the following is obtained,
\begin{equation}
\frac{A_m}{B_m}=\frac{\mathrm{K}_m(ka)}{\mathrm{I}_m(ka)}.\label{eq:r1}
\end{equation}
Using Ohm's law, $\mathbf{J}_s=\sigma_\mathrm{2D}\mathbf{E}_t$, where $\mathbf{J}_s$, $\sigma_\mathrm{2D}$, and $\mathbf{E}_t$ are surface current density, two-dimensional surface conductivity, and tangential electric field, respectively, and exploiting current continuity equation, the following relation for the surface charge is derived,
\begin{eqnarray}
\rho_s &=&\frac{\sigma_{2D}(\omega_{k,m})}{i\omega_{k,m}}A_m\mathrm{I}_m(ka)\left(\frac{m^2}{a^2}+k^2\right)\nonumber\\
&&\times \exp i(kz+m\varphi-\omega_{k,m}t)+\mathrm{c.c.}, \label{eq:surfcharge}
\end{eqnarray}
where $\mathrm{c.c.}$ denotes complex conjugate of previous terms. By using Eq.~(\ref{eq:surfcharge}) and substituting in perpendicular electric field boundary condition, another relation for coefficients is derived,
\begin{equation}
\frac{B_m}{A_m}=\frac{\epsilon_0\epsilon_1k\mathrm{I}'_m(ka)-\frac{\sigma_{2D}(\omega_{k,m})}{i\omega_{k,m}}\left(\frac{m^2}{a^2}+k^2\right)\mathrm{I}_m(ka)}{\epsilon_0\epsilon_2k\mathrm{K}'_m(ka)}.\label{eq:r2}
\end{equation}
In the above relation, $\epsilon_0$ is vacuum permittivity and primes denote derivation with respect to the argument. Combining Eq.~(\ref{eq:r1}) and Eq.~(\ref{eq:r2}) leads to the dispersion relation of plasmons,
\begin{equation}
\frac{\sigma_{2D}(\omega_{k,m})}{i\omega_{k,m}\epsilon_0a}=\frac{\epsilon_1\frac{\mathrm{I}'_m(ka)}{\mathrm{I}_m(ka)}-\epsilon_2\frac{\mathrm{K}'_m(ka)}{\mathrm{K}_m(ka)}}{\left[m^2+(ka)^2\right]}\cdot ka. \label{eq:disp}
\end{equation}
SPs correspond to $k\rightarrow\infty$ part of dispersion relation. So the following result is gained by using large argument approximation of modified Bessel functions \cite{abramowitz1964handbook},
\begin{equation}
\sigma_{2D}(\omega(k))=\frac{2i\omega(k)\epsilon_0\bar{\epsilon}}{k}, \label{eq:dispapp}
\end{equation}
where $\bar{\epsilon}$ is the arithmetic average of $\epsilon_1$ and $\epsilon_2$. In the above relation we recast $\omega_{k,m}$ to $\omega(k)$, for two reasons: First, the index $k$ is continuous and it is convenient to consider it as a variable rather than an index, merely and second, we omit the $m$ index, because there is no dependence on $m$ in right hand side of Eq.~(\ref{eq:dispapp}).  Until now no assumption for surface conductivity profile is used. So the derived dispersion relation is general for any arbitrary profile of surface conductivity, and is not restricted to graphene. An interesting result, obtained from Eq.~(\ref{eq:dispapp}), is that for large $k$'s all the modes become degenerate and also dispersion relation does not depend on cylinder radius. This is an important result, which one could expect previously, because for large enough radius and wavenumber, SPs are mostly confined to the graphene and so Indeed does not sense cylinder radius anymore and the results should approach to extended graphene ones.  For illustrating this assertion, the dispersion relation is derived by assuming Drude approximation, Eq.~(\ref{eq:drude}), of graphene conductivity and neglecting any damping;
\begin{equation}
\omega_m(k)=\sqrt{\frac{e^2E_F}{\pi\hbar^2\epsilon_0a}}\frac{1}{\sqrt{\theta_m(ka)}},\label{eq:dispgen}
\end{equation}
where
\begin{equation}
\theta_m(x)\equiv x\frac{\epsilon_1\frac{\mathrm{I}'_m(x)}{\mathrm{I}_m(x)}-\epsilon_2\frac{\mathrm{K}'_m(x)}{\mathrm{K}_m(x)}}{\left(m^2+x^2\right)}.
\end{equation}
If we further assume $\epsilon_1=\epsilon_2\equiv\bar{\epsilon}$, then the result is more simplified,
\begin{equation}
\omega_{k,m}^2=\frac{e^2E_F}{\pi\hbar^2a\epsilon_0\bar{\epsilon}}g_m(x),
\end{equation}
where
\begin{equation}
g_m(x)=(m^2+x^2)\mathrm{I}_m(x)\mathrm{K}_m(x).
\end{equation}
In the derivation of above relation, the Wronskian property of modified Bessel functions, $\mathrm{I}'_m(x)\mathrm{K}_m(x)-\mathrm{I}_m(x)\mathrm{K}'_m(x)=1/x$, is utilized \cite{abramowitz1964handbook}. Fig.~\ref{fig:disp}   shows dispersion curves, Eq.~(\ref{eq:dispgen}), for some lower order modes. In drawing this figure, it is assumed that $a=100\,\mathrm{nm}$ and $E_F=0.4\,\mathrm{eV}$. Fig.~\ref{fig:disp}(a) depicts $\hbar\omega$ normalized  to Fermi energy versus $k$ normalized to Fermi wavenumber $k_F=E_F/\hbar v_F$, where $v_F=10^6\,m/s$ is Fermi velocity, and Fig.~\ref{fig:disp}(b) draws plasmon energy versus $ka$. In this figure, it can be seen that for $k\gtrsim 0.2k_F$ or equivalently $\hbar\omega\gtrsim 0.1E_F$, all the modes become degenerate, as an evidence for the previous statement. The asymptotic form of dispersion relation, assuming Drude-like conductivity for graphene, is:
\begin{equation}
\omega(k)=\sqrt{\frac{e^2E_F}{2\pi\hbar^2\epsilon_0\bar{\epsilon}}}\cdot\sqrt{k}.\label{eq:grspdisp}
\end{equation}
This result resembles the extended graphene case and confirms the previously mentioned assertion. It must be noticed that for this formulation to be valid, $E_F$ must lie in one of the following two regions,
\label{eq:efrange}
\begin{equation}
0.1<E_F<\frac{0.08\pi\epsilon_0\bar{\epsilon}}{ek},\label{eq:efrange1}
\end{equation}
\begin{equation}
\frac{ek}{8\pi\epsilon_0\bar{\epsilon}}<E_F<0.1.\label{eq:efrange2}
\end{equation}
In the above intervals, $E_F$ is in $\mathrm{eV}$ unit. This restriction is due to the previously mentioned range of the validity of Drude approximation, used in this formulation.

\begin{figure}[tb]
\centering
\includegraphics[width=.45\linewidth]{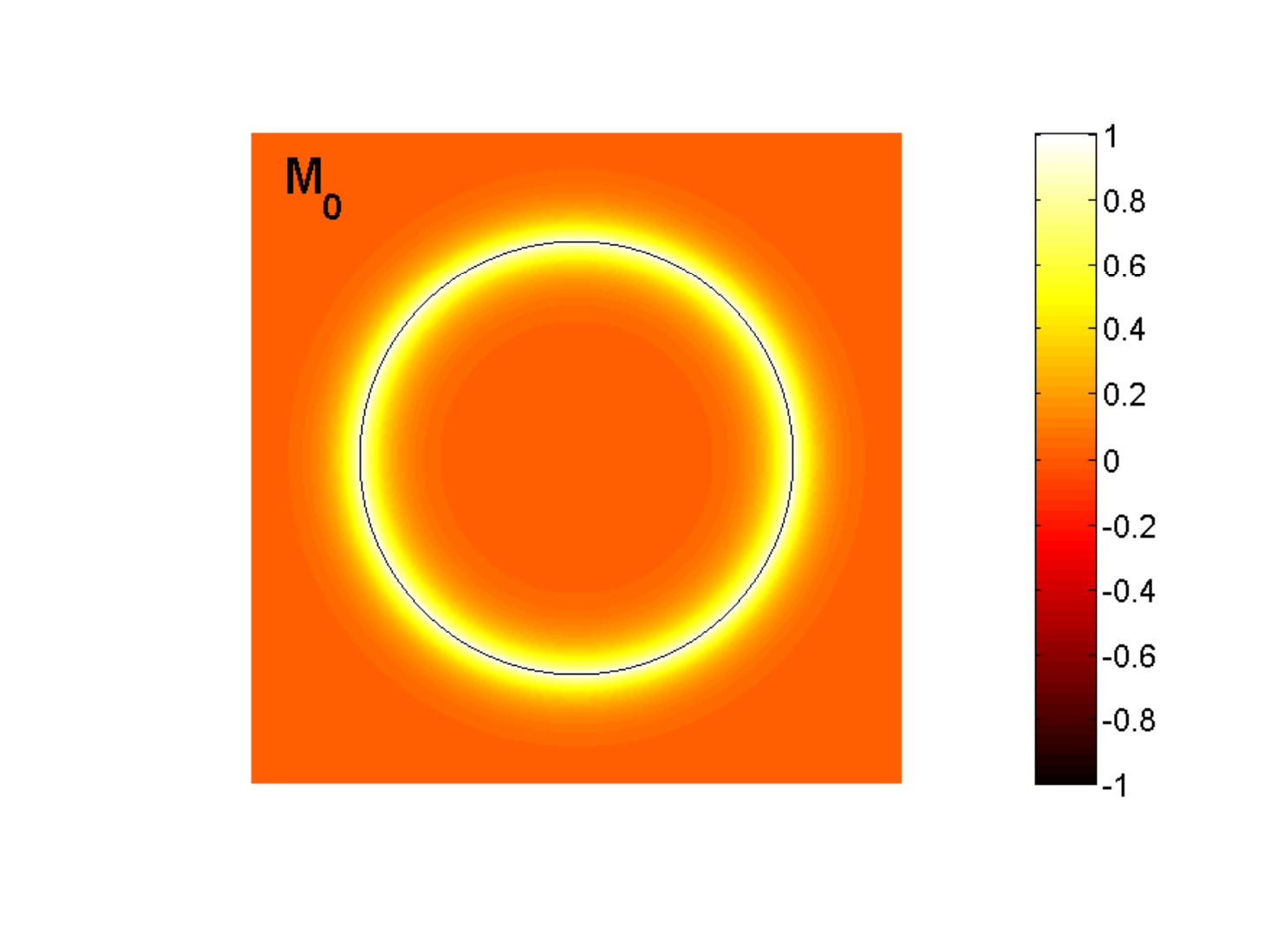}
\includegraphics[width=.45\linewidth]{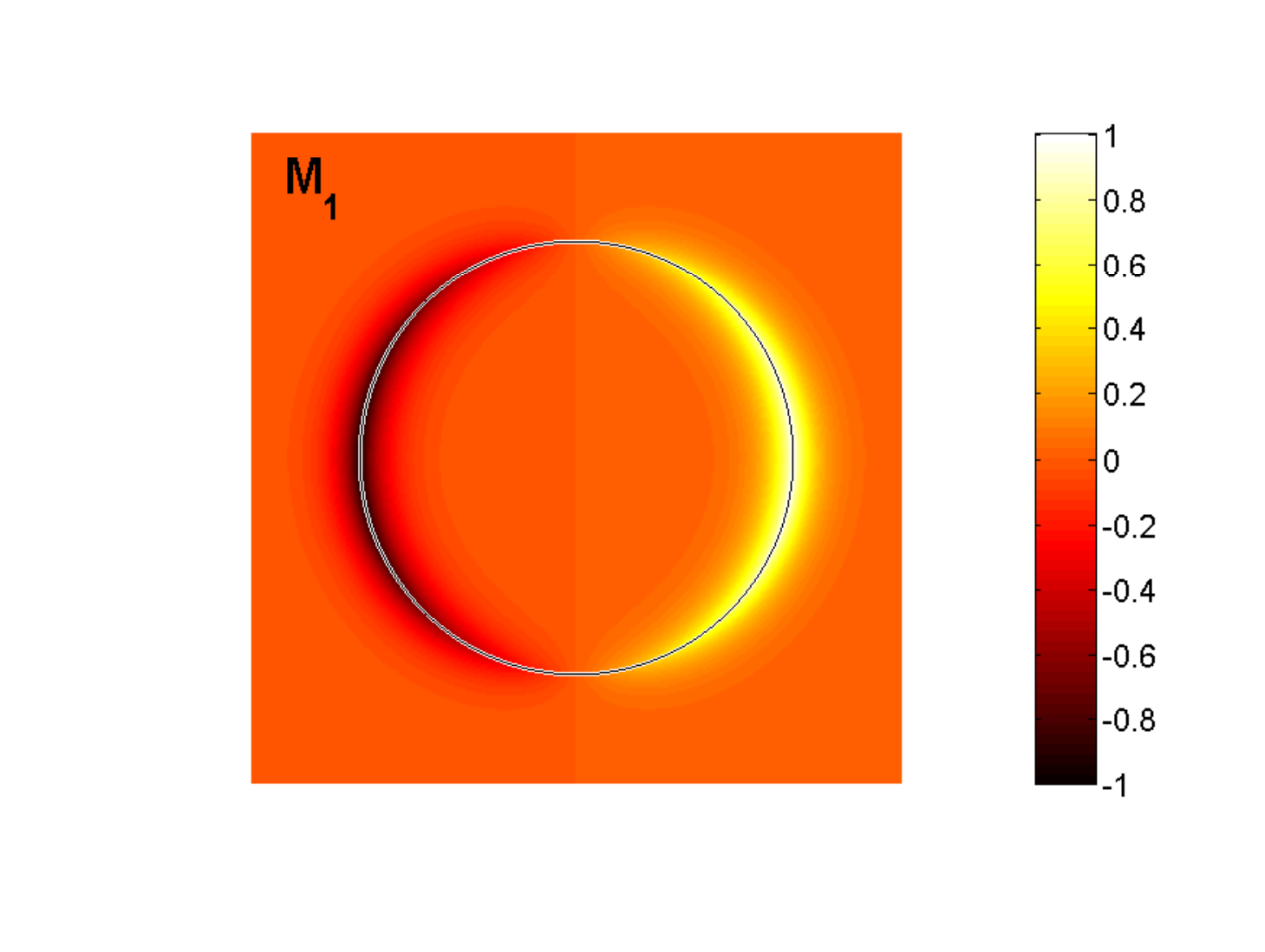}\\
\includegraphics[width=.45\linewidth]{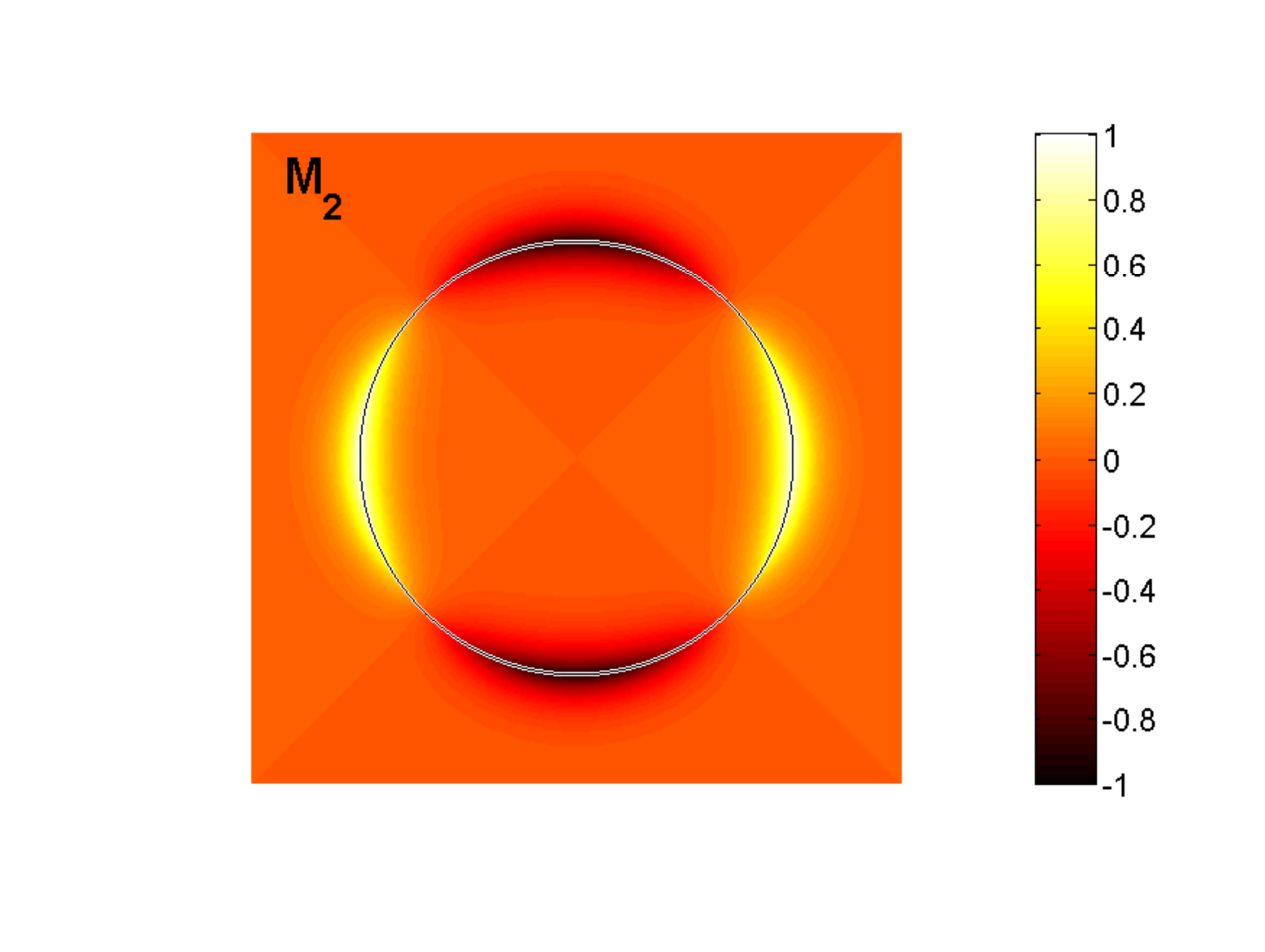}
\includegraphics[width=.45\linewidth]{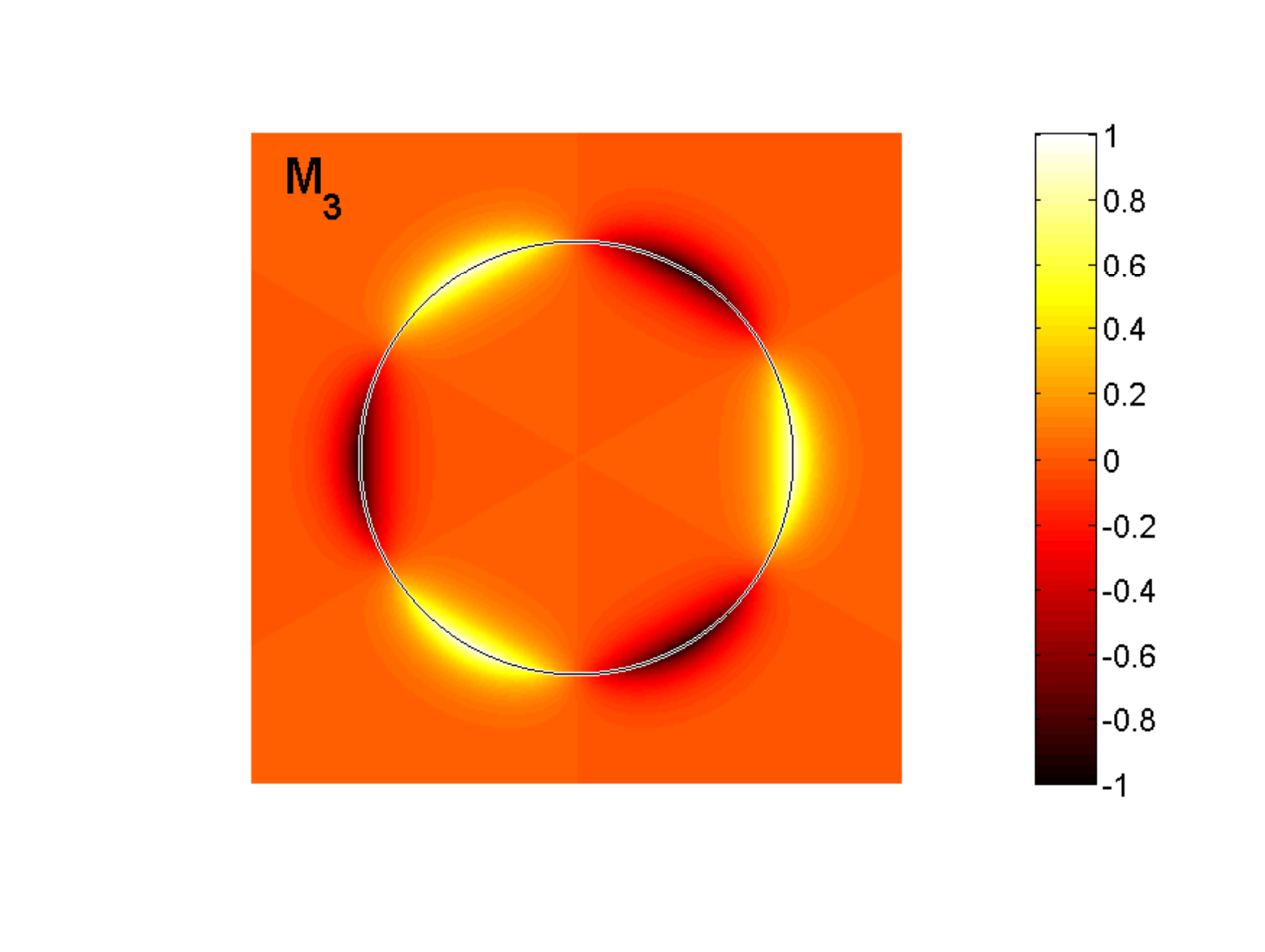}
\caption{(Color online) Cross section view of the first four normalized potential modes so that their maximums become unity. Notation $\mathrm{M}_m$ is used for modes, where $m$ is the mode index defined in the paper.}
\label{fig:phimodes}
\end{figure}

Finally, the normalized potential can be written in the following form,
\begin{equation}
\phi(\mathbf{r},t)=\left\lbrace\begin{array}{ll}
\frac{\mathrm{I}_m(k\rho)}{\mathrm{I}_m(ka)}\exp i(kz+m\varphi-\omega_{k,m}t) &\rho\le a\\
\frac{\mathrm{K}_m(k\rho)}{\mathrm{K}_m(ka)}\exp i(kz+m\varphi-\omega_{k,m}t) &\rho>a.
\end{array}\right.
\end{equation}
The normalized potential profiles of first four lower order modes are shown in Fig.~\ref{fig:phimodes}.

\begin{figure}[tb]
\centering
\includegraphics[width=\linewidth]{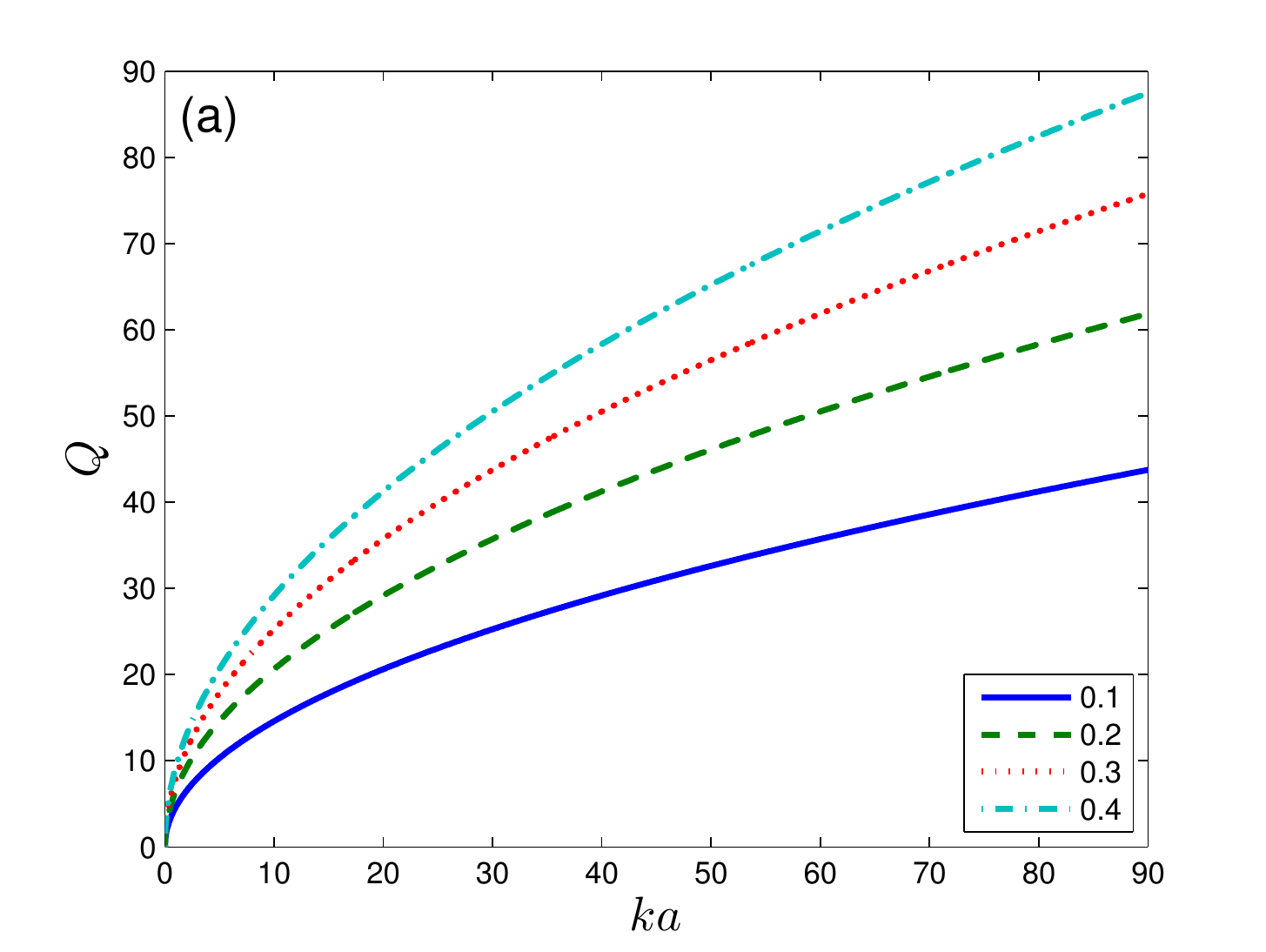}
\includegraphics[width=\linewidth]{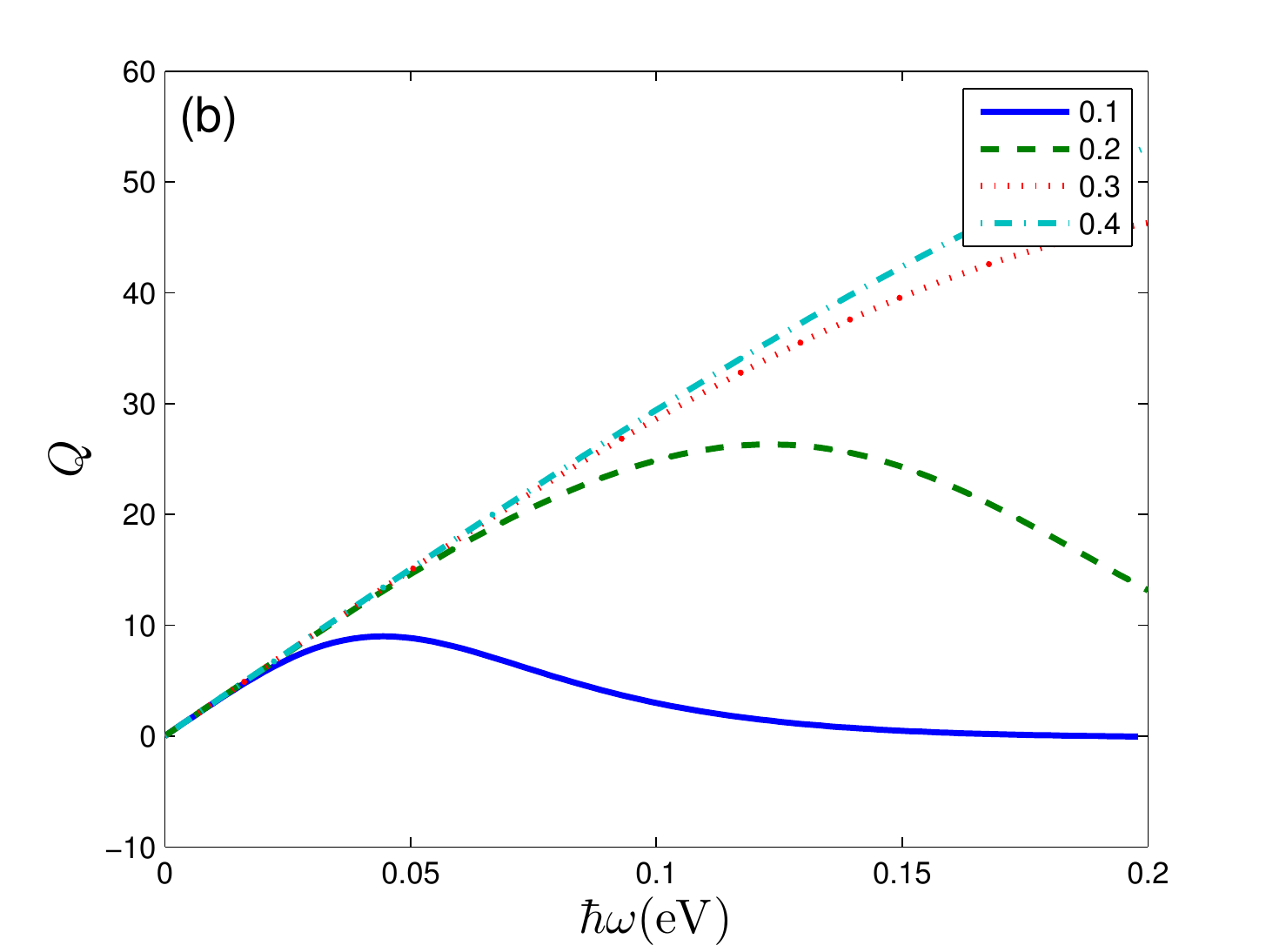}
\caption{(Color online) The Quality factor of modes as a function of (a) wavenumber and (b) frequency for different Fermi energies. In figure (b) one can see the range which the linear approximation is valid.}
\label{fig:Q}
\end{figure}

The other important parameter, which will be encountered later in this work, is the Quality factor of modes. If we write $\Omega(k)=\omega(k)-i\gamma(k)$ and $\sigma_{2D}=\sigma'_{2D}+i\sigma''_{2D}$, and replace $\omega(k)$ by $\Omega(k)$ in Eq.~(\ref{eq:dispapp}), by equating real and imaginary parts of both sides, and assuming $\gamma\ll\omega$, which is valid for the frequency range which will be used, the following two equations are obtained,
\begin{eqnarray}
\sigma'_{2D}(\omega(k)) &=& \frac{2\epsilon_0\bar{\epsilon}\gamma(k)}{k},\label{eq:sig1}\\
\sigma''_{2D}(\omega(k)) &=& \frac{2\epsilon_0\bar{\epsilon}\omega(k)}{k}.\label{eq:sig2}
\end{eqnarray}
After dividing Eq.~(\ref{eq:sig2}) by Eq.~(\ref{eq:sig1}), the Quality factor of SP modes is,
\begin{equation}
Q(\omega(k))=\frac{\sigma''_{2D}(\omega(k))}{2\sigma'_{2D}(\omega(k))}\simeq\frac{\omega(k)}{2\tau^{-1}},\label{eq:Q}
\end{equation}
where $\omega(k)$ is calculated from Eq.~(\ref{eq:grspdisp}) or Eq.~(\ref{eq:sig2}). Fig.~\ref{fig:Q}  shows the Quality factor of SP modes as a function of plasmon wavenumber and energy for different values of $E_F$ using Eq.~(\ref{eq:Q}). Regarding Fig.~\ref{fig:Q}(b), It can be seen that as $E_F$ increases the linear approximation of $Q$ becomes more accurate.

\subsection{\label{SPquant}Quantization of SP Hamiltonian}
In the electrostatic regime, the SP Hamiltonian is decomposed of kinetic and potential parts, \mbox{$H_\mathrm{sp}=H_\mathrm{kin}+H_\mathrm{pot}$}, where for the potential part,
\begin{equation}
H_\mathrm{pot}=\frac{1}{2}\int_{S_g}\rho_s\phi\,\mathrm{d}^2r, \label{eq:hpot1}
\end{equation}
such that $\rho_s$ is the surface charge density of graphene due to the existence of plasmons and $\phi$ is total electric potential. The integration runs over the graphene's surface,  $S_g$. The kinetic part is \cite{arista2001interaction}:
\begin{equation}
H_\mathrm{kin}=\frac{1}{2}n_{s0}m^*_e\int_{S_g}\left|\mathbf{v}_e\right|^2\,\mathrm{d}^2r, \label{eq:hkin1}
\end{equation}
where $n_{s0}$, $m^*_e$, and $\mathbf{v}_e$ are the surface density of electrons in equilibrium, a suggested plasmonic electron effective mass, not equal to common electron effective mass and average drift velocity of electrons, respectively. The above kinetic Hamiltonian resembles that of 3D electron gas one. Indeed, we suggest using of the same formulation for graphene, but with a modified electron effective mass. We propose to find the effective mass by equating the Drude conductivity of graphene to the 3D electron gas one \cite{ashcroft1976solid},
\begin{equation}
\sigma(\omega)=\frac{i\epsilon_0\omega_p^2}{\omega+i\gamma},\label{eq:sigmadr}
\end{equation}
where plasma frequency is defined by $\omega_p=e^2n_0/\epsilon_0m^*_e$ and $n_0$ is electron number density. By equating Eq.~(\ref{eq:sigmadr}) with graphene conductivity, Eq.~(\ref{eq:drude}), one can find $m^*_e=n_{s0}\pi\hbar^2/E_F$.

For the purpose of quantizing the plasmon field, we write all the field variables in the Hamiltonian, Eq.~(\ref{eq:hpot1}) and Eq.~(\ref{eq:hkin1}), as a linear combination of plasmon modes. So the electric potential can be written in the following form,
\begin{equation}
\phi(\mathbf{r},t)=\sum_{k,m}C_{k,m}\phi_{k,m}(\rho)\exp i(kz+m\varphi-\omega_{k,m}t)+\mathrm{c.c.},\label{eq:phi}
\end{equation}
where
\begin{eqnarray}
\phi_{k,m}(\rho) &=&\phi_{k,m}^+(\rho)+\phi_{k,m}^-(\rho),\\
\phi_{k,m}^-(\rho) &=&\Theta(-\rho+a)\frac{\mathrm{I}_m(k\rho)}{\mathrm{I}_m(ka)},\\
\phi_{k,m}^+(\rho) &=&\Theta(\rho-a)\frac{\mathrm{K}_m(k\rho)}{\mathrm{K}_m(ka)},
\end{eqnarray}
and $C_{k,m}$'s are expansion coefficients. Using this potential, surface charge density can be derived exploiting perpendicular electric field boundary condition,
\begin{eqnarray}
\rho_s&=&\sum_{k,m}C_{k,m}\frac{\sigma_{2D}(\omega_{k,m})}{i\omega_{k,m}}\left(\frac{m^2}{a^2}+k^2\right)\nonumber\\
&\times&\exp i(kz+m\varphi-\omega_{k,m}t)+\mathrm{c.c.}\label{eq:rhos}
\end{eqnarray}
The only remaining quantity is drift velocity. The drift velocity can be derived using Newton's second law, $-e\mathbf{E}=m^*_e\mathrm{d}\mathbf{v}_e/\mathrm{d}t$, where $\mathbf{E}=-\nabla\phi$ is electric field. After some algebra the following result is obtained,
\begin{eqnarray}
\mathbf{v}_e(\mathbf{r}_\parallel,t)&=&\frac{e}{m_e^*}\sum_{k,m}\left[\frac{m}{a}\hat{\varphi}+k\hat{\mathbf{z}}\right]\frac{1}{\omega_{k,m}}C_{k,m}\nonumber \\
&\times&\exp i(kz+m\varphi-\omega_{k,m}t)+\mathrm{c.c.},\label{eq:ve}
\end{eqnarray}
where $\mathbf{r}_\parallel$ is the in-plane position vector. By substituting Eq.~(\ref{eq:rhos}) and Eq.~(\ref{eq:ve}) into Hamiltonians, Eq.~(\ref{eq:hpot1}) and Eq.~(\ref{eq:hkin1}), and after some cumbersome algebra,  the following results are found,
\begin{eqnarray}
&&\hspace{-1.3cm}H_\mathrm{pot}=\frac{A_g}{2}\sum_{k,m}\frac{\sigma_{2D}(\omega_{k,m})}{i\omega_{k,m}}\left(\frac{m^2}{a^2}+k^2\right)\times\nonumber\\
&&\hspace{-1.1cm}\left[C_{k,m}C_{-k,-m}\exp i(\omega_{k,m}+\omega_{-k,-m})t+C_{k,m}C_{k,m}^*\right]\nonumber\\
&&\hspace{-1.1cm}+\mathrm{\mathrm{c.c.}},\label{eq:hpot}\\
&&\hspace{-1.3cm}H_\mathrm{kin}=\frac{A_ge^2n_{s0}}{2m_e^*}\sum_{k,m}\left(\frac{m^2}{a^2}+k^2\right)\times\nonumber\\
&&\hspace{-1.1cm}\left[\frac{-C_{k,m}C_{-k,-m}}{\omega_{k,m}\omega_{-k,-m}}\exp i(\omega_{k,m}+\omega_{-k,-m})t+\frac{C_{k,m}C_{k,m}^*}{\omega_{k,m}^2}\right]\nonumber\\
&&\hspace{-1.1cm}+\mathrm{c.c.}\label{eq:hkin}
\end{eqnarray}
Throughout the paper, we assume that $A_g$ and $L$ are the hypothetical area and length of graphene cylinder, respectively, and $k$ and $m$ run over all possible index values. In deriving the above relations the following orthogonality properties are exploited,
\begin{eqnarray}
\int_{-L/2}^{L/2}e^{i(k-k')z}\,\mathrm{d} z &=&L\delta_{k,k'},\\
\int_0^{2\pi}e^{i(m-m')\varphi}\,\mathrm{d}\varphi &=&2\pi\delta_{m,m'},
\end{eqnarray}
where $\delta$ represents Kronecker delta function. By combining Eq.~(\ref{eq:hpot}) and Eq.~(\ref{eq:hkin}) and assuming \mbox{$\omega_{k,m}=\omega_{-k,-m}$} the SP Hamiltonian is obtained,
\begin{eqnarray}
H_\mathrm{sp}&=&\frac{A_g}{2}\sum_{k,m}\frac{\sigma_{2D}(\omega_{k,m})}{i\omega_{k,m}}\left(\frac{m^2}{a^2}+k^2\right)\nonumber\\
&\times&\left(C_{k,m}C_{k,m}^*+C_{k,m}^*C_{k,m}\right).
\end{eqnarray}
The above Hamiltonian is analogous to harmonic oscillator's one, such that by the following substitution and assuming negligible damping, $H_\mathrm{sp}$ recasts to the operator form,
\begin{eqnarray}
C_{k,m} &\rightarrow&\gamma_{k,m}(\omega_{k,m})\hat{a}_{k,m},\\
C^*_{k,m} &\rightarrow&\gamma_{k,m}(\omega_{k,m})\hat{a}^\dagger_{k,m},
\end{eqnarray}
where $\gamma_{k,m}$ is defined as:
\begin{equation}
\gamma_{k,m}(\omega_{k,m})=\left(\frac{\hbar\omega_{k,m}^2}{A_g\left|\sigma''_{2D}\right|(\omega_{k,m})}(m^2/a^2+k^2)\right)^{1/2}.\label{eq:gammakm}
\end{equation}
Using these relations, the SP Hamiltonian is simplified in the following operator form,
\begin{eqnarray}
\hat{H}_\mathrm{sp}&=&\sum_{k,m}\frac{\hbar\omega_{k,m}}{2}\left(\hat{a}^\dagger_{k,m}\hat{a}_{k,m}^{\phantom{\dagger}}+\hat{a}_{k,m}^{\phantom{\dagger}}\hat{a}^\dagger_{k,m}\right)\nonumber\\
&=&\sum_{k,m}\hbar\omega_{k,m}\left(\hat{a}^\dagger_{k,m}\hat{a}_{k,m}^{\phantom{\dagger}}+\frac{1}{2}\right),
\end{eqnarray}
where $\hat{a}_{km}$ and $\hat{a}_{km}^\dagger$ are annihilation and creation operators of an SP in the mode $k,m$, respectively, and obey bosonic algebra \cite{scully1997quantum},
\begin{eqnarray}
\lbrack\hat{a}_{k,m},\hat{a}_{k',m'}^\dagger] &=&\delta_{k,k'}\delta_{m,m'},\\
\lbrack\hat{a}_{k,m},\hat{a}_{k',m'}] &=& 0,\\
\lbrack\hat{a}^\dagger_{k,m},\hat{a}_{k',m'}^\dagger] &=& 0.
\end{eqnarray}
By substituting Eq.~(\ref{eq:gammakm}) in Eq.~(\ref{eq:phi}), the electric field operator is obtained,
\begin{eqnarray}
\hspace{-.5cm}\hat{\mathbf{E}}(\mathbf{r},t)&=&\sum_{k,m}\gamma_{k,m}(\omega_{k,m})\nonumber\\
&\times&\left[\mathbf{M}_{k,m}(\mathbf{r})\hat{a}_{k,m}(t)+\mathbf{M}_{k,m}^*(\mathbf{r})\hat{a}_{k,m}^\dagger(t)\right]\!,\label{eq:elecf}
\end{eqnarray}
where
\begin{equation}
\mathbf{M}_{k,m}(\mathbf{r})=\phi'_{k,m}(\rho)\hat{\rho}+\frac{im}{\rho}\phi_{k,m}(\rho)\hat{\varphi}+ik\phi_{k,m}(\rho)\hat{\mathbf{z}}.\label{eq:M}
\end{equation}
In Eq.~(\ref{eq:elecf}), we switched to the Heisenberg picture, where the time dependence is completely transferred to the operators.

\section{\label{sec:gainHam}Active Medium Hamiltonian}
\begin{figure}[tb]
\centering
\includegraphics[width=.45\linewidth]{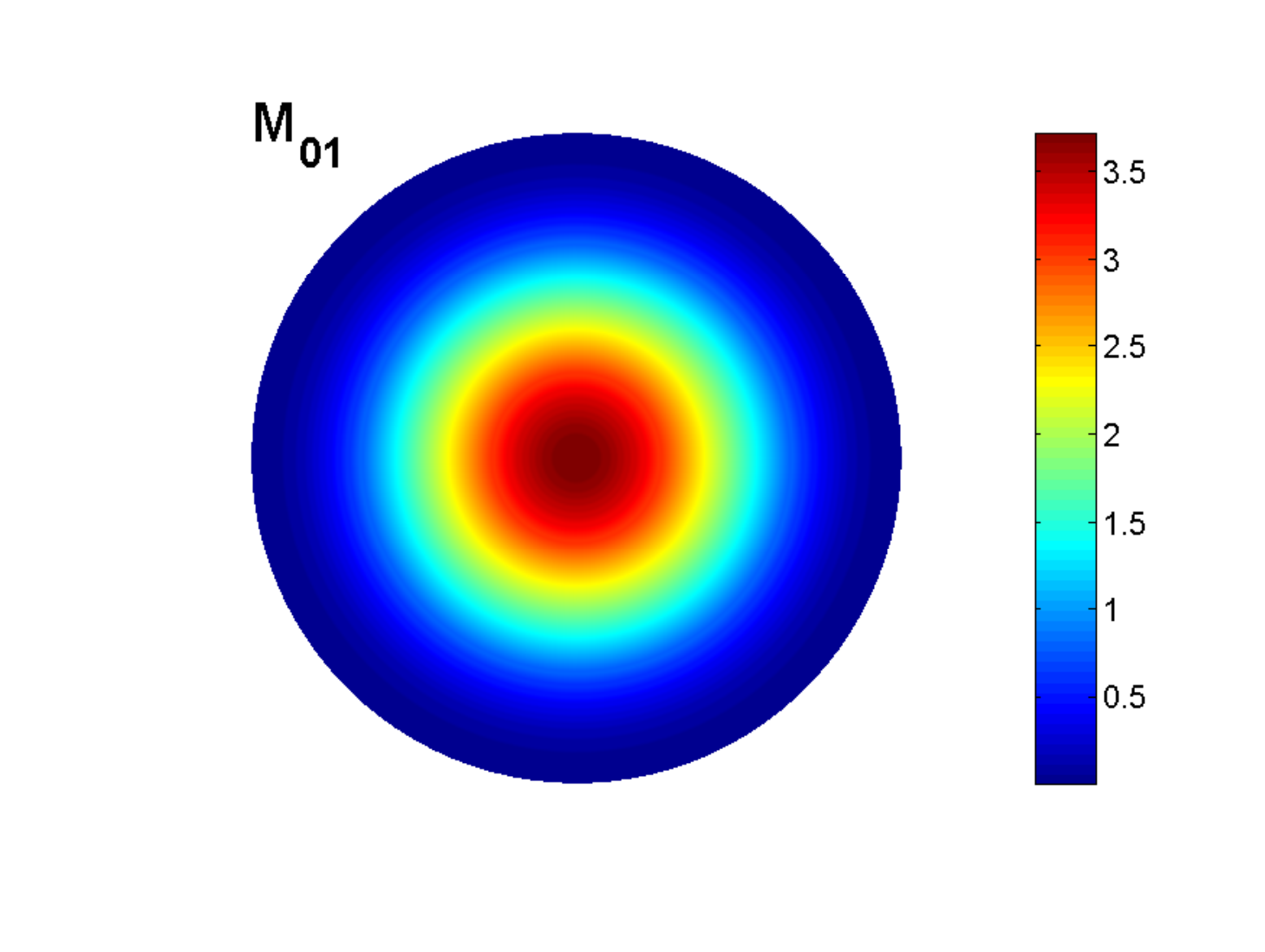}
\includegraphics[width=.45\linewidth]{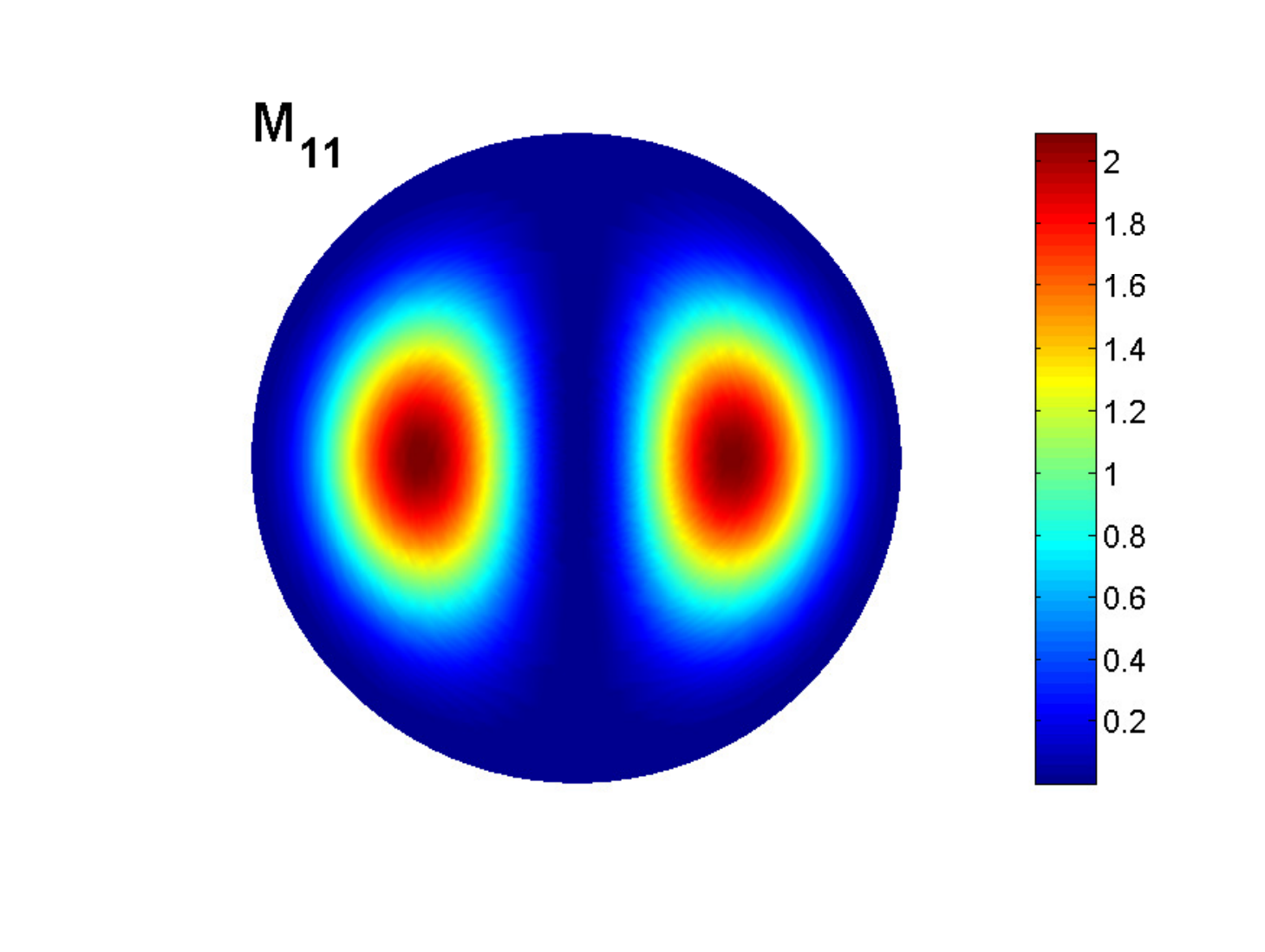}\\
\includegraphics[width=.45\linewidth]{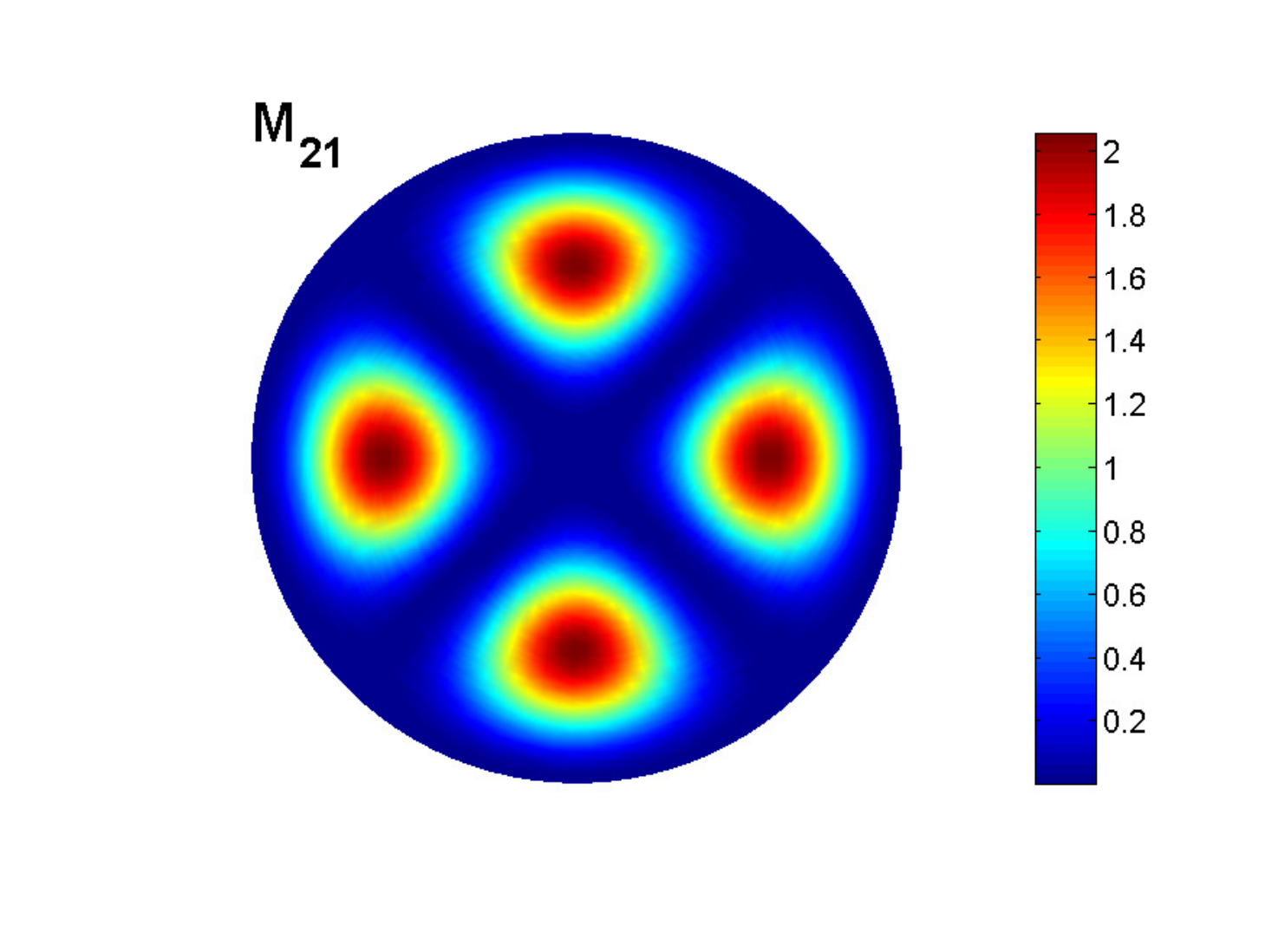}
\includegraphics[width=.45\linewidth]{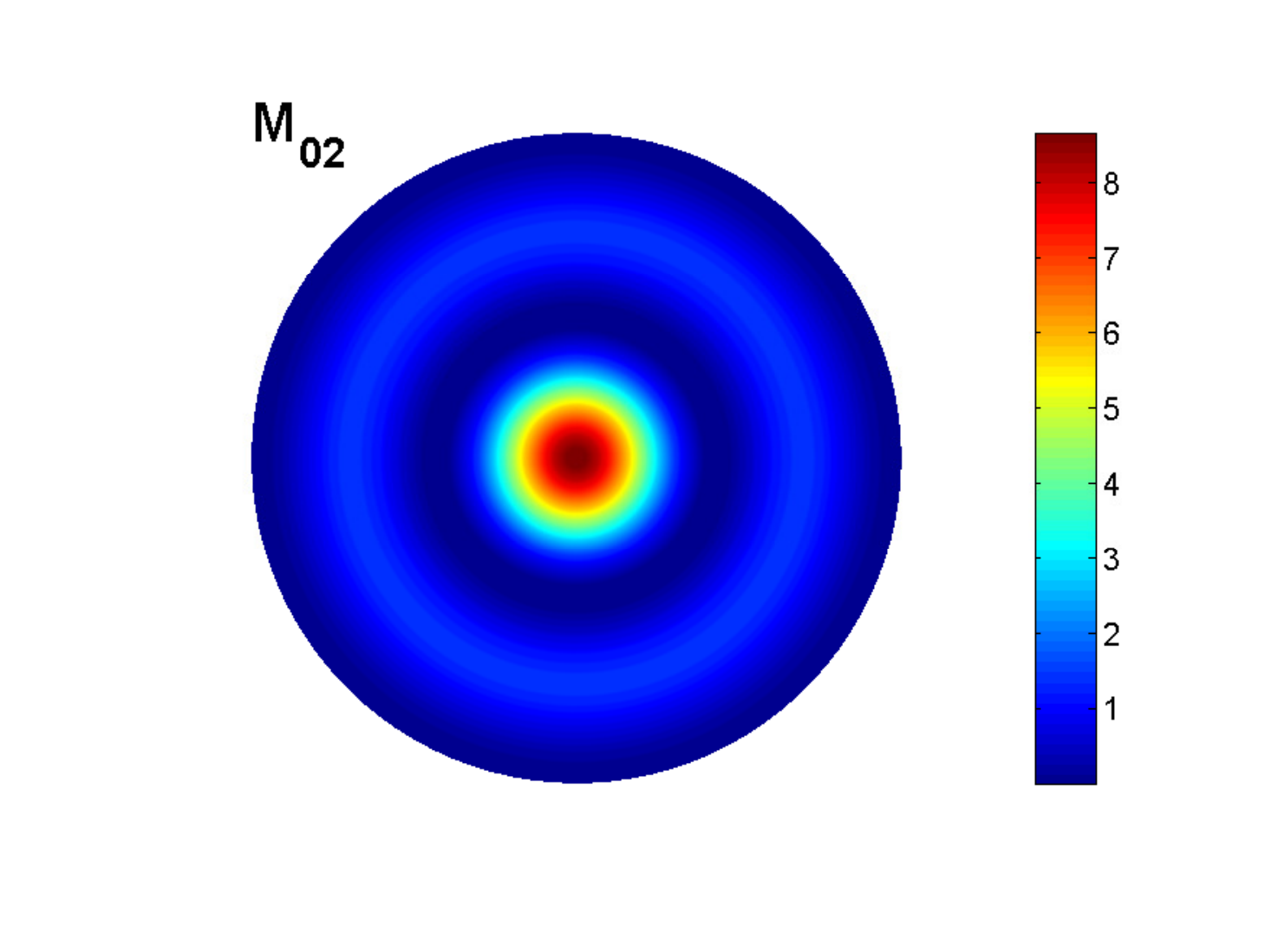}
\caption{(Color online) Cross section view of the absolute moduli squared of the first four lower order eigenfunctions. The notation $\mathrm{M}_{nl}$ shows the mode with quantum number $nl$ discussed in the paper.}
\label{fig:wrmodes}
\end{figure}

In the present work, we propose to utilize a QW of radius $b$ and volume $V_\mathrm{QW}$ as the gain medium. For further investigation of the structure, energy levels and wavefunctions should be derived, using the well known Schr\"{o}dinger equation. For the sake of simplicity and obtaining a rule of thumb, appropriate for design purposes, the infinite wall boundary condition is applied. The wavefunctions of such a structure are:
\begin{equation}
\psi_{nl}(\rho,\phi,z)=\left\lbrace\begin{array}{ll}
\frac{\exp i(k_z z+n\phi)}{\sqrt{V_\mathrm{QW}} \mathrm{J}_{n+1}(x_{nl})}\mathrm{J}_n(\frac{x_{nl}}{b}\rho) & \rho\le b\\
0 & \rho>b,
\end{array}\right.
\end{equation}
where $\psi_{nl}$, $x_{nl}$, and $k_z$ are $nl$'th eigenfunction, the $l$'th zero of $n$'th order Bessel function of the first kind, and wavenumber along the longitudinal direction, respectively. The first four lower order modes are sketched in Fig.~\ref{fig:wrmodes}. The eigenenergy associated with $nl$'th mode is $E=E_{nl}+E_c$, where $E_c=\frac{\hbar^2k_z^2}{2m^*_w}$ and energy levels $E_{nl}$ are:
\begin{equation}
E_{nl}=\frac{\hbar^2x_{nl}^2}{2m^*_wb^2},
\end{equation}
where $m^*_w$ is the electron's effective mass in the wire.

Maximum coupling between the QW and graphene is achieved when both are in resonance with each other, i.e. $\hbar\omega_\mathrm{sp}=E_\mathrm{e}-E_\mathrm{g}$, where $\omega_\mathrm{sp}$, $E_\mathrm{e}$ and $E_\mathrm{g}$ are SP's angular frequency, and excited and ground states energy, respectively. Thus, for design purposes, $b$ maybe chosen such that the energy difference between excited and ground state coincides with plasmon energy. After some substitution and rearranging, the following result is found for quantum wire's radius,
\begin{equation}
b=\sqrt{\frac{\hbar}{2m_w^*\omega_{sp}}(x_{n_el_e}^2-x_{n_gl_g}^2)}.\label{eq:qwradius}
\end{equation}

Another important quantity, which has a vital role in the next section, is the dipole moment. The value of dipole moment represents how much the coupling strength is. It can be shown that for our structure, dipole moments only have nonzero values between states which have the same quantum number $n$. In addition, it is simple to show that dipole moment has only radial component. dipole moment between $nl$ and $nl'$ states is:
\begin{equation}
\mathbf{d}_{nlnl'}=2\pi ebf_{nll'}\hat{\rho},
\end{equation}
where
\begin{equation}
f_{nll'}=\frac{1}{\mathrm{J}_{n+1}(x_{nl})\mathrm{J}_{n+1}(x_{nl'})}\int_0^1\rho^2\mathrm{J}_n(x_{nl}\rho)\mathrm{J}_{n}(x_{nl'}\rho)\,\mathrm{d}\rho,
\end{equation}
is a dimensionless number, which depends on $n$, $l$, and $l'$ and it is independent of $b$. If $\psi_{01}$ and $\psi_{02}$ are considered as the states which are in resonance with a specific plasmon mode then this number is equal to $0.09722$.

According to spectral decomposition theorem \cite{hassani2013mathematical}, the active medium Hamiltonian, in the basis which diagonalizes itself, can be written in the following form,
\begin{equation}
\hat{H}_\mathrm{am}=\sum_i E_i\hat{\sigma}_{ii},
\end{equation}
where $i$ is a representative of all the discrete and continuous quantum numbers and runs over all the possible states and $\hat{\sigma}_{ii}=\left|i\right>\!\left<i\right|$. If we assume that the only transition, strongly coupled to the plasmon field, is $p\rightarrow q$,  and further, seting the zero level of energy to the halfway between these two states, then the active medium Hamiltonian can be written in the simple form,
\begin{equation}
\hat{H}_\mathrm{am}=\frac{\hbar\omega_{qp}}{2}\hat{\sigma}_z,
\end{equation}
where the following definitions are used,
\begin{eqnarray}
&&\hat{\sigma}_z =\left|q\right>\!\left<q\right|-\left|p\right>\!\left<p\right|,\\
&&\hbar\omega_{qp}=E_q-E_p.
\end{eqnarray}

\section{\label{sec:intHam}Interaction Hamiltonian and Spasing}
We assume that the active medium can be approximated as a dipole. Accuracy of this approximation depends on how large the multipole terms are, relative to dipole term, in the potential multipole expansion. As a rule of thumb, the more the distance between quantum wire and graphene, the more accurate results are obtained. By using this assumption, the interaction Hamiltonian can be writen as $\hat{H}_\mathrm{int}=-\hat{\mathbf{d}}\cdot\hat{\mathbf{E}}$, where $\hat{\mathbf{d}}$ and $\hat{\mathbf{E}}$ are dipole moment and electric field operators, respectively. $\hat{\mathbf{d}}$ can be written down in the following form \cite{scully1997quantum},
\begin{equation}
\hat{\mathbf{d}}=\mathbf{d}_{qp}\left(\hat{\sigma}_++\hat{\sigma}_-\right),
\end{equation}
where $\mathbf{d}_{qp}$ is a dipole matrix element, associated with $p\rightarrow q$ transition, and $\hat{\sigma}_+=\left|q\right>\!\left<p\right|$ and $\hat{\sigma}_-=\left|p\right>\!\left<q\right|$ are rising and lowering ladder operators, respectively. After using these relations and considering energy conservation, the interaction Hamiltonian can be written as
\begin{equation}
\hat{H}_\mathrm{int}=-\hbar\left[\Omega_{kmqp}(\mathbf{r}_0)\hat{\sigma}_+\hat{a}+\Omega_{kmqp}^*(\mathbf{r}_0)\hat{a}^\dagger\hat{\sigma}_-\right],
\end{equation}
where $\mathbf{r}_0$ is position vector of dipole and Rabi frequency, $\Omega_{kmqp}=-\hat{\mathbf{d}}\cdot\hat{\mathbf{E}}/\hbar$ \cite{scully1997quantum}, is written as
\begin{equation}
\Omega_{kmqp}(\mathbf{r}_0)=\frac{\gamma_{k,m}(\omega_{k,m})}{\hbar}\mathbf{d}_{qp}\cdot\mathbf{M}_{k,m}(\mathbf{r}_0).\label{eq:rabi1}
\end{equation}
Taking into account that $\mathbf{d}_{qp}$ has only a component along the radial direction and substituting Eq.~(\ref{eq:M}) into Eq.~(\ref{eq:rabi1}), the following result is obtained,
\begin{equation}
\Omega_{kmqp}(\mathbf{r}_0)=\left\lbrace\begin{array}{ll}
\frac{\gamma_{k,m}(\omega_{k,m})}{\hbar}d_{qp}k\frac{\mathrm{I}'_m(k\rho_0)}{\mathrm{I}_m(ka)} &\rho_0<a\\
\frac{\gamma_{k,m}(\omega_{k,m})}{\hbar}d_{qp}k\frac{\mathrm{K}'_m(k\rho_0)}{\mathrm{K}_m(ka)} &\rho_0>a,
\end{array}\right.\label{eq:rabi}
\end{equation}
where $\rho_0$ is quantum wire radial position.

\begin{figure}[tb]
\centering
\includegraphics[width=\linewidth]{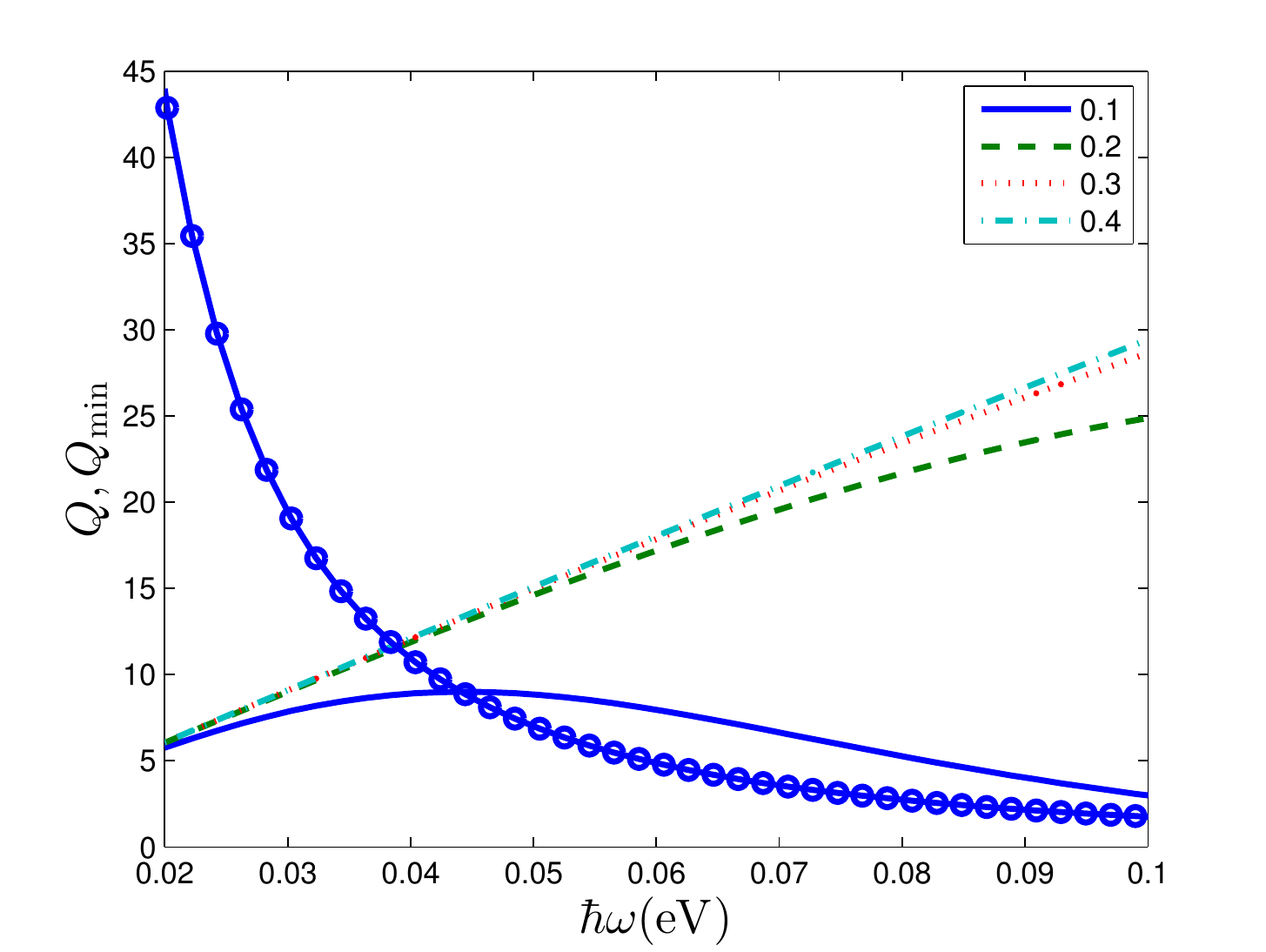}
\caption{(Color online) The solid, dashed, dotted, and dash-dotted lines show the Quality factor of modes versus plasmon energy for different values of Fermi level. Circle marked line represents minimum required value of Quality factor. The intersection of modes' Quality factor curves with minimum Quality factor one determines threshold frequency which is the minimum allowable frequency for spasing.}
\label{fig:Qmin}
\end{figure}

The spasing condition can be written as \cite{apalkov2014proposed,stockman2010spaser},
\begin{equation}
\frac{(\gamma'_{km}+\Gamma_{qp})^2}{(\gamma'_{km}+\Gamma_{qp})^2+(\omega_{qp}-\omega_{k,m})^2}\sum_{k_z}\left|\Omega_{kmqp}\right|^2\geq \gamma'_{km}\Gamma_{qp},\label{eq:spscond}
\end{equation}
where $\Gamma_{qp}$, and $\gamma'_{km}$ are the damping rate of polarization and plasmon mode $k,m$, respectively, and $k_z$ runs over all possible transverse wavenumbers. By substituting the rabi frequency, Eq.~(\ref{eq:rabi}), into spasing condition, Eq.~(\ref{eq:spscond}), and assuming near resonance region, after some manipulations, we find that for spasing to be able to occur the Quality factor of SP modes should be higher than $Q_\mathrm{min}$,
\begin{equation}
Q_\mathrm{min}=\frac{2\pi^2a\hbar\epsilon_0\Gamma_{qp}\mathrm{Z}_m^2(ka)}{\left|d_{qp}\right|^2k_Fk\mathrm{Z}_m^{'2}(k\rho_0)}\cdot\left[\epsilon_1\frac{\mathrm{I}'_m(ka)}{\mathrm{I}_m(ka)}-\epsilon_2\frac{\mathrm{K}'_m(ka)}{\mathrm{K}_m(ka)}\right],\label{eq:Qcondg1}
\end{equation}
or
\begin{equation}
Q_\mathrm{min}=\frac{\pi ae^2v_F\Gamma_{qp}\mathrm{Z}_m^2(ka)}{\left|d_{qp}\right|^2\omega_{k,m}^2\mathrm{Z}_m^{'2}(k\rho_0)}\cdot\left[\epsilon_1\frac{\mathrm{I}'_m(ka)}{\mathrm{I}_m(ka)}-\epsilon_2\frac{\mathrm{K}'_m(ka)}{\mathrm{K}_m(ka)}\right],\label{eq:Qcondg2}
\end{equation}
where
\begin{equation}
\mathrm{Z}_m(x)=\left\lbrace\begin{array}{ll}
\mathrm{I}_m(x) &\qquad \rho_0<a\\
\mathrm{K}_m(x) &\qquad\rho_0>a.
\end{array}\right.
\end{equation}
For large values of $k$, that is the case for SPs, the condition for minimum Quality factor reduces to
\begin{equation}
Q_\mathrm{min}\simeq\frac{4\pi^2a\hbar\epsilon_0\bar{\epsilon}\Gamma_{qp}}{\left|d_{qp}\right|^2k_Fk}, \label{eq:Qcondk}
\end{equation}
or as a function of angular frequency,
\begin{equation}
Q_\mathrm{min}\simeq\frac{2\pi ae^2v_F\Gamma_{qp}}{\left|d_{qp}\right|^2\omega^2(k)}=\frac{av_F\Gamma_{qp}}{2\pi f_{nll'}^2b^2\omega^2(k)}.\label{eq:Qcondw}
\end{equation}
Fig.~\ref{fig:Qmin}
illustrates, graphically, which frequency regions are allowable for spasing. The figure shows that for a fixed $E_F$, there can exist an intersection point between $Q$ and $Q_\mathrm{min}$ curves, which from now on we call it threshold frequency and denote by $\omega_\mathrm{th}$. Spasing can occur for frequencies higher than $\omega_\mathrm{th}$. This dependency can be driven analytically by combining Eq.~(\ref{eq:Qcondw}) and Eq.~(\ref{fig:Q}),
\begin{equation}
\omega_\mathrm{th}=\frac{av_F\Gamma_{qp}}{\pi b^2f_{nll'}^2\tau}.\label{eq:wth}
\end{equation}
It can be seen that threshold frequency depends on the ratio of the cylinder radius and cross section area of the quantum wire.

After this long discussion, we have arrived at the point, which we can design a graphene cylinder based spaser by using derived formulas. We want to propose a design procedure:\\
\begin{enumerate}
\item For a given $\omega_\mathrm{sp}$ calculate $b$ from Eq.~(\ref{eq:qwradius}),
\item Calculate $a$ using Eq.~(\ref{eq:wth}) so that $\omega_\mathrm{sp}>\omega_\mathrm{th}$ or equivalently \[a<\frac{\pi b^2f_{nll'}^2\tau}{v_F\Gamma_{qp}}\cdot\omega_\mathrm{sp}^3,\]
\item Determine $E_Fk$ using Eq.~(\ref{eq:grspdisp}),
\item Derive $E_F$ and $k$ separately so that the validation range, Eq.~(\ref{eq:efrange}), is satisfied.
\end{enumerate}
There exists a freedom for assigning $E_F$ and $k$ values, separately, as long as the validation range is fulfilled. Increasing $k$ confines SPs more and more, so $E_F$ can be utilized for changing spot size.

As an example, we design a spaser for $\hbar\omega_\mathrm{sp}=0.1\,\mathrm{eV}$. The calculated $b$ for this specific frequency is $15.1\,\mathrm{nm}$, the maximum value for $a$ is $1.73\,\mu\mathrm{m}$ and $E_Fk=4.33\times10^7$. The maximum value for $E_Fk$ is $17.39\times10^7$ provided that we choose $E_F>0.1\,\mathrm{eV}$, Eq.~(\ref{eq:efrange1}). As long as $E_F>0.1\,\mathrm{eV}$, we can change $E_F$ to focus SPs beam. The smallest spot is obtained where $E_F=0.1\,\mathrm{eV}$. In drawing Fig.~\ref{fig:Qmin}, we use these values for $a$ and $b$, and further assume \mbox{$\Gamma_{qp}=3.6\,\mathrm{meV}$} \cite{apalkov2014proposed}.
  
\section{Conclusion}
In summary, we have suggested a structure for spasing. The structure has been analyzed theoretically using full quantum mechanical approach which treats both the field and matter quantum mechanically. For quantizing the SP field and in writing the kinetic energy of electrons inside graphene, a special effective mass has been defined. The spasing condition for the structure has been derived by quantizing the Hamiltonian of the system. Finally, a design procedure has been proposed and a spaser for plasmon energy of $\hbar\omega_\mathrm{sp}=0.1\,\mathrm{eV}$ has been designed. Throughout the paper, the electrostatic approximation was used. In this regime, the polarization aspects of wave and the effects due to them e.g. distinguishing between TE and TM modes could not be investigated. For applications, which the wave polarization is important, the full wave method must be used.
\bibliography{references}

\end{document}